\documentclass[structabstract]{aa}

\usepackage{graphicx}
\usepackage{color}
\usepackage{epsfig}
\usepackage[varg]{txfonts}
\usepackage{natbib}
\usepackage[utf8]{inputenc}
\bibpunct{(}{)}{;}{a}{}{,}

\def\kms {\rm{km~s^{-1}}}

\def\Mpc {\rm Mpc}

\def\AA {\r{A}}

\begin{document}

   \title{The BL-Lac gamma-ray blazar PKS 0447-439 \\
as a probable member of a group of galaxies \\
at $z=0.343$}

\titlerunning{The BL-Lac gamma-ray blazar PKS 0447-439}

   \author{H. Muriel \inst{1,2}; C. Donzelli \inst{1,2}; A.C. Rovero \inst{3} \and A. Pichel \inst{3}  }

\institute{Instituto de Astronom\'ia
  Te\'orica y Experimental (IATE, CONICET-UNC), Laprida 854, X5000BGR, C\'ordoba, Argentina
         \and Observatorio Astron\'omico, Universidad Nacional de C\'ordoba, Laprida 854, 
          X5000BGR, C\'ordoba, Argentina \and Instituto de Astronom\'ia y F\'isica del Espacio 
(IAFE, CONICET-UBA), Av. Inte. G\"uiraldes 2620, C1428ZAA, Buenos Aires, Argentina\\
         \email{hernan@oac.unc.edu.ar}
}

\date{Received ... ; accepted ...}


\abstract
   {The BL-Lac blazar PKS 0447-439 is one of the brightest HE gamma-ray sources that were first detected by Fermi-LAT. It 
was also detected by H.E.S.S. at VHE gamma-rays, which allowed constraining the redshift of PKS 0447-439 by 
considering the attenuation caused by gamma-ray interactions with ambient photons in the extragalactic background light (EBL). This constraint 
agreed with color-magnitude and spectroscopic redshift constraints ($0.179 < z < 0.56$), Recently, however,  
a much higher redshift was proposed for this blazar ($z > 1.2$). This value was debated 
because if true, it would imply either that the relevant absorption processes of gamma-rays are not well understood or that 
the EBL is dramatically different from what is believed today. This high redshift was not confirmed by three 
independent new spectroscopic observations at high signal-to-noise ratios. The scenario is clear evidence of 
the difficulties in estimating the redshift of BL-Lac objects, whose optical spectra are typically featureless. 
Neither of the estimated redshifts for PKS 0447-439 are confirmed
as yet.}
   {Given that BL-Lac are typically hosted by elliptical galaxies, which in turn are associated with groups, 
we aim to find the host group of galaxies of PKS 0447-439. The ultimate goal is to estimate a redshift 
for this blazar.}
   {Spectra of twenty-one objects in the field of view of PKS 0447-439 were obtained with the Gemini 
Multi-Object Spectrograph. Based on the redshifts and coordinates of these galaxies, we searched for groups of galaxies. Using a deep catalog of groups, we studied the probability of finding
by chance a
group of galaxies in the line of sight of PKS 0447-439.}
   {We identified a group of galaxies that was not previously cataloged at $z = 0.343$ with seven members, a virial 
radius of $0.42$ $\Mpc,$ and a velocity dispersion of 622 km s$^{-1}$. We found that the probability of
the host galaxy of PKS 0447-439 to be a member of the new group  
is $\gtrsim 97\%$. Therefore, we propose to adopt $z = 0.343 \pm 0.002$ as the most likely redshift for 
PKS 0447-439.}
{}


\maketitle
%

\section{Introduction}
\label{sec:intro}

Blazars are a subclass of active galactic nuclei (AGN) where the relativistic jet points close to 
the observer line of sight (e.g., \citealt{Urry:1995}). They are one of the most powerful astrophysical 
sources and are characterized by rapid, strong, and irregular flux variability (e.g., \citealt{Falomo:2014}). 
There are two populations of blazars, flat-spectrum radio quasars (FSRQ), and BL Lacertae (BL-Lac) objects. 
The latter are classified as low-, intermediate-, and high-peaked (LBL, IBL, and HBL) BL-Lacs, depending 
on the position of the lower energy bump in the spectral energy distribution (SED) \citep{Fossati:1998}. 
If the blazar evolution sequence is correct \citep{Bottcher:2002, Cavaliere:2002}, FSRQs and BL-Lac 
objects would have an evolutionary link. FSRQs would be the earliest phase of an evolving sequence 
with a final long phase at the HBL stage, where the gas accretion rate (and jet density) from the 
host galaxy decreases. While the evolution of blazars is a matter of debate, there is some evidence 
supporting this transition from FSQR to BL-Lac \citep{Giommi:2012, Ajello:2014}.

The SED of BL-Lacs are dominated by strong nonthermal emission from the jet, 
which covers up all the characteristic spectral features of the host galaxy and usually prevents determining spectroscopic redshifts. Given this difficulty, other methods have been proposed for 
redshift determination, particularly important for the luminosity function at the gamma-ray domain 
where BL-Lacs are the most frequent extragalactic objects (e.g., \citealt{Reimer:2013}).
Gamma-ray radiation in the range of very high-energy (VHE; E$>$100 GeV) is strongly attenuated by the 
photon-photon interaction with the extragalactic background light (EBL) and for the more energetic 
gamma-rays, with the cosmic microwave background (CMB). As a consequence, all discovered extragalactic 
VHE sources (mostly BL-Lac blazars) are relatively close ($z <0.6$, except perhaps for the recent not 
yet confirmed detection of the blazar S3 0218+357 at $z=0.944$ -- ATel \#6349). At these redshifts, the 
attenuation due to photon-photon interaction for high-energy (HE; E$>$100 MeV) gamma-rays is negligible. 
Then, for BL-Lac blazars gamma-ray astronomy provides an alternative method for estimating the redshift, 
which is to model the drop in the SED from HE to VHE caused by the photon-photon interaction 
(e.g., \citealt{Dwek:2013}). To do this, a distribution of the not very well-known EBL field has to 
be assumed, and both the spectrum at HE and VHE have to be measured. Thus, by extending the HE 
spectrum to higher energies, the drop HE-to-VHE is estimated and related to a true redshift value 
\citep{Prandini:2010}. A caution for this procedure has been given by arguing that such an extension 
may not be the actual connection in the SED of blazars \citep{Costamante:2012}.

It has also been proposed that the absorption features in the far-ultraviolet spectra of BL-Lac 
objects can be used to estimate limits to the redshift of blazars \citep{Danforth:2010, Danforth:2013, 
Furniss:2013}. The far-ultraviolet spectra of blazars show a smooth 
continuum with intergalactic-medium absorption features (the Ly$\alpha$ forest) that can be used to set 
lower limits to the blazar redshift. The upper limit is set by statistically treating the non-detection 
of absorbers beyond a certain redshift.

The BL-Lac blazar PKS 0447-439 is one of the brightest HE sources that were first detected by Fermi-LAT 
\citep{Abdo:2009}. It was also detected at VHE by H.E.S.S. with a soft derived spectrum (photon 
index 3.89) and no indication of break or curvature \citep{Abramowski:2013}. Modeling the drop in the 
SED from HE to VHE, and assuming the EBL density of \citet{Franceschini:2008}, it was possible to 
derive an estimate of redshift for PKS 0447-439 of $z\sim0.2$ \citep{Prandini:2012}. This was 
based on H.E.S.S. preliminary results at VHE \citep{Zech:2011} and assuming average spectral 
properties of blazars. A more certain upper limit of $z<0.59$ for the redshift of PKS 0447-439 was 
found by \citet{Abramowski:2013} using a lower limit model of the EBL and assuming there are no 
upturns in the flux above the HE band; they have also found that a satisfactory description of the 
SED would not be feasible if the redshift upper limit is different
from $z\lesssim0.4$.
These results agree with the spectroscopic value $z=0.205$ reported in 1998 using data 
taken with the CTIO 4 m telescope, in which the identification of weak Ca II absorption lines was 
claimed \citep{Perlman:1998}. A lower limit of $z>0.176$ for the redshift of PKS 0447-439 was also 
estimated from the optical V magnitude \citep{Landt:2008}. 
On the other hand, the distribution of known redshifts for BL-Lac gamma-ray blazars has a peak 
approximately in the range $z\approx0.1-0.3$ with a tail to higher values ($z\lesssim1.5$) 
\citep{Ackermann:2011}. Then, if the evolutionary link of the blazar sequence mentioned above is correct, 
the peak of this redshift distribution would mostly be populated by the blazar type HBL, as this is 
suggested to be the last long phase of the sequence. Provided PKS 0447-439 is classified as HBL 
(e.g., \citealt{Abramowski:2013}), it would be reasonable to adopt a value for the redshift of PKS 0447-439 
close to or within the peak.

While all these estimations showed a coherent picture for the distance of PKS 0447-439, a significantly 
higher lower limit of $z>1.246$ for the redshift of this blazar was announced based on the 
identification of the  Mg II $\lambda$2800 line doublet in absorption \citep{Landt:2012}. This estimate was 
considered very controversial as a high redshift like this for a VHE source would imply either that 
the relevant absorption processes of gamma-rays are not well understood or that the EBL is 
dramatically different from what is believed today. An alternative less exotic explanation for this 
high redshift was proposed by arguing that the distant TeV blazar emission could be compatible with 
secondary photons produced by energetic protons from the blazar jet propagating over almost rectilinear cosmological 
distances \citep{Aharonian:2013}.

Immediately after the claim of \citet{Landt:2012} was announced, several spectroscopic observations were 
performed trying to confirm this high redshift \citep{Pita:2012, Fumagalli:2012, Rovero:2013}. They 
all confirmed the absorption line at 6280 \AA, which could result in a high-redshift 
value if identified as Mg II $\lambda$2796.82. However, they associated this line with a known telluric 
absorption line, invalidating the claim that this is a very distant blazar. No other spectral features 
were detected, therefore no spectroscopic redshift for PKS 0447-439 was established from any of these observations. 
It is worth to mention that the observations performed by \citet{Rovero:2013} were made with the Gemini 
8.1 m telescope at a significantly high signal-to-noise ratio (S/N $\sim$200 at 4000 \AA ~to $\sim$500 at 
6000 \AA), the highest of all spectroscopic observations of PKS 0447-439. Despite this, 
\citet{Rovero:2013} were not able to identify the Ca II absorption lines used by \citet{Perlman:1998} 
to report a redshift of 0.205, which makes the latter an unconfirmed spectroscopic redshift. The fact 
that these Ca II lines were not seen by recent high-quality observations could be attributed to either 
a significant flux variability of AGNs, or to a poor line identification by \citet{Perlman:1998}, 
which is probably the case because they identified very weak Ca II lines on an otherwise 
featureless optical spectrum.

Since BL-Lacs are typically hosted by elliptical galaxies, it is expected that these objects are associated with
groups or clusters of galaxies. \citet{Fried:1993} found an excess in the galaxy density around 
BL-Lacs at low and intermediate redshifts. Other studies confirmed that BL-Lacs inhabit
group environments \citep{Wurtz:1993, Falomo:1993, Pesce:1994, Smith:1995, Pesce:1995, Wurtz:1997}.
Motivated by this evidence, and considering the controversy on the 
redshift of PKS 0447-439, we have used spectroscopic data taken on this AGN and twenty-one other objects 
in the field to study its environment. Data taken on PKS 0447-439 were analyzed previously \citep{Rovero:2013}. 
The aim of this work is to spectroscopically analyze
a sample
of 21 objects in the field of view of PKS 0447-439 to find their host group of galaxies.  
The structure of this paper is as follows: 
Section \ref{sec:Obs} describes the photometric and spectroscopic observations. 
In Sect. \ref{sec:Res} we present the analysis and results, and in 
Sect. \ref{sec:Con} we summarize our conclusions.

Where necessary, a cosmology with $H_{0} = 70$ $\kms$  $Mpc^{-1}$, 
${\Omega }_{m} = 0.25$, and 
${\Omega }_{\Lambda} = 0.75$ is applied.

\section{Observations}
\label{sec:Obs}

Spectra for PKS 0447-439 and twenty-one other objects in the field were obtained 
with the Gemini Multi-Object Spectrograph (GMOS), program GS-2012B-Q25 (PI: A.C. Rovero).
We centered the field on PKS 0447-439 covering a region of $\sim5\times5$ arcmin$^2$. 
A multislit mask was created using a pre-image provided by Gemini.
This image consisted of $1\times30$ s exposure in Sloan $g$ (G0325) filter
with a scale of 0.1456 arcsec pixel$^{-1}$.
Additional objects were chosen considering that concentration limits the
total number of target fitting on the GMOS mask.
The mask was designed by the slit-positioning algorithm (SPA) that
determines which objects will be placed on the mask. 
Object selection by the SPA
is based on object priority and position on the frame, ensuring that
there is at least a two-pixel separation between object slits.
About 30-40 slits can typically be placed on the mask, depending on the
slit width. In our case, the final mask consisted of 22 slits.
Figure \ref{fig:gemini} shows the objects selected for spectroscopy.

The pre-image was taken on 28 September 2012, while the spectroscopic data were
acquired in queue mode on 21 November 2012 using the designed
multislit mask. Slitlet dimensions were 1 arcsec by 4 arcsec. The
grating in use was B600$+_{-}$G5323, which has a ruling density of
600 lines/mm. Three exposures of 900 s each were obtained with central
wavelengths of 497 nm, 502 nm, and 507 nm to remove the gaps
between the CCD chips. Science targets have thus a total exposure time
of 0.75 hours. The seeing was $\sim 0.9$ arcsec during the
observations. Spectra typically cover the wavelength range 3900-5600
$\AA$, although because the wavelength range depends on the slit
position, some spectra start at $\sim 3600$ $\AA$ while others end at
$\sim 7000$ $\AA$. Flatfields, spectra of the
standard star $LTT$ $7787$, and the copper-argon $CuAr$ lamp were also
acquired to perform flux and wavelength calibrations.
A binning of $2\times2$ was used, yielding a scale of 0.1456 arcsec
per pixel and a theoretical dispersion of
$\sim 0.9$ \AA~ per pixel. Observed objects together with their
coordinates and $g$ magnitudes are listed in Table \ref{tab:data}.

\subsection{Photometry at Bosque Alegre}
\label{sec:bosque}

Additionally, we procured images at B, V, R, and I from Johnson system 
using the 1.54 m telescope at Bosque Alegre, C\'ordoba. Images were taken on
14 September 2014 together with Landolt standards \citep{Landolt:2007}. The telescope
has attached an Apogee Alta U9 CCD camera at the Newtonian focus, which gives an
scale of 27.5 arcsec mm$^{-1}$. The camera was binned $3\times3,$ resulting in a 0.74 
arcsec pixel. Exposure times were 150, 100, 50, and 80 s in the B, V, R, and I
filters, respectively. The apparent magnitudes for PKS 0447-439 were then measured to be
$m_B = 14.55$, $m_V = 13.99$, $m_R = 13.58$, and $m_I = 14.35$.
Unfortunately, the other objects targeted for spectroscopy are fainter than
the limiting magnitude of the CCD camera.

\subsection{Data reduction}
\label{sec:Data}

All science and calibration files were retrieved from the Gemini
Science Archive hosted by the Canadian Astronomy
Data Center. The data reduction described below was carried out with
the Gemini IRAF package. Flatfields were
derived with the task {\it gsflat} and the flatfield exposures.
Spectra were reduced using {\it gsreduce,} which
performs a standard data reduction (i.e., bias, overscan, and
cosmic rays removal) and applies
the flatfield derived with {\it gsflat}. GMOS-South detectors are read
with three amplifiers that generate files with three distinct extensions. The task {\it gmosaic} was used to
generate data files with a single extension.
The sky level was removed interactively using the task {\it gskysub,}
and the spectra were extracted using
{\it gsextract}.

Flux calibration was performed using the spectra of the standard star
$LTT$ $7787$, acquired with an identical
instrument configuration. Spectra of $CuAr$ lamps were obtained
immediately after science target observations and were used to achieve wavelength calibration using the
task {\it gswavelength}. The sensitivity
function of the instrument was derived using {\it gsstandard} and the
reference file for $LTT$ $7787$ provided
by Gemini Observatory. Science spectra were flux calibrated with {\it
gscalibrate,} which uses the
sensitivity function derived by {\it gsstandard}.

Images were pre-processed through the standard Gemini pipeline that
corrects bias, dark current, and flat field.
Recent photometric zero points that yield standard magnitudes were
taken from the Gemini website.
The foreground galactic extinction around PKS 0447-439 is $A_{g} = 0.045$ mag
(from the NASA Extragalactic Database).

Aperture photometry for the observed objects was made using {\it DAOPHOT}
within {\it IRAF}. Total magnitudes were
obtained through a series of circular diaphragms until the total flux
converged. Previous to these
calculations we made a careful background subtraction
considering an area with an inner radius of
5" and outer radius of 10" centered on each object. In addition,
neighbor objects were properly masked to avoid
flux contamination.

Redshifts for the targeted objects around PKS 0447-439 were calculated
using the {\it FXCOR} routine within {\it IRAF}.
This task computes radial velocities by deriving the Fourier cross
correlation between two spectra.
In our case, the spectra obtained for the globular clusters and planetary
nebulae in NGC 7793 were used as a template.
All these spectra were taken with the same GMOS configuration during a
previous Gemini run (GS-2011B-Q10, PI: C. Donzelli). These spectra have
the necessary absorption or emission lines to correlate
with those of the sample objects.
In general terms, we have computed redshifts using four or more lines,
usually $H\gamma$, [OII] (3727 \AA), $H\beta$, [OIII] (5007 \AA), and
the Ca II H+K absorption lines.

   \begin{figure*}
   \centering
   \includegraphics[width=10cm]{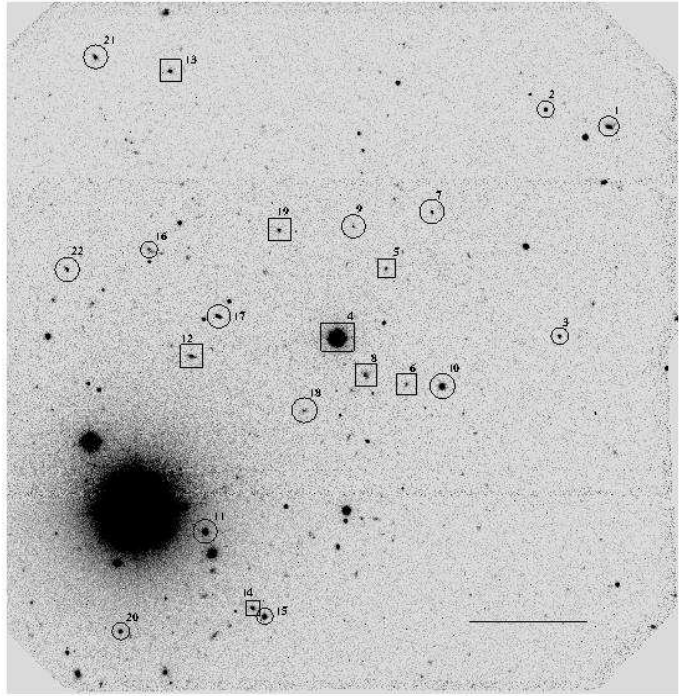}
   \caption{GMOS $g'$-band image centered on PKS 0447-439. North is at top and east to the left. 
Observed objects are marked with circles and squares (group members) and labeled according to 
the slit number. The lower right line indicates 1 $arcmin$.}
   \label{fig:gemini}
   \end{figure*}

\section{Results}
\label{sec:Res}


\subsection{Group of galaxies at $z=0.343$}

\begin{figure}
   \centering
   \includegraphics[width=8cm]{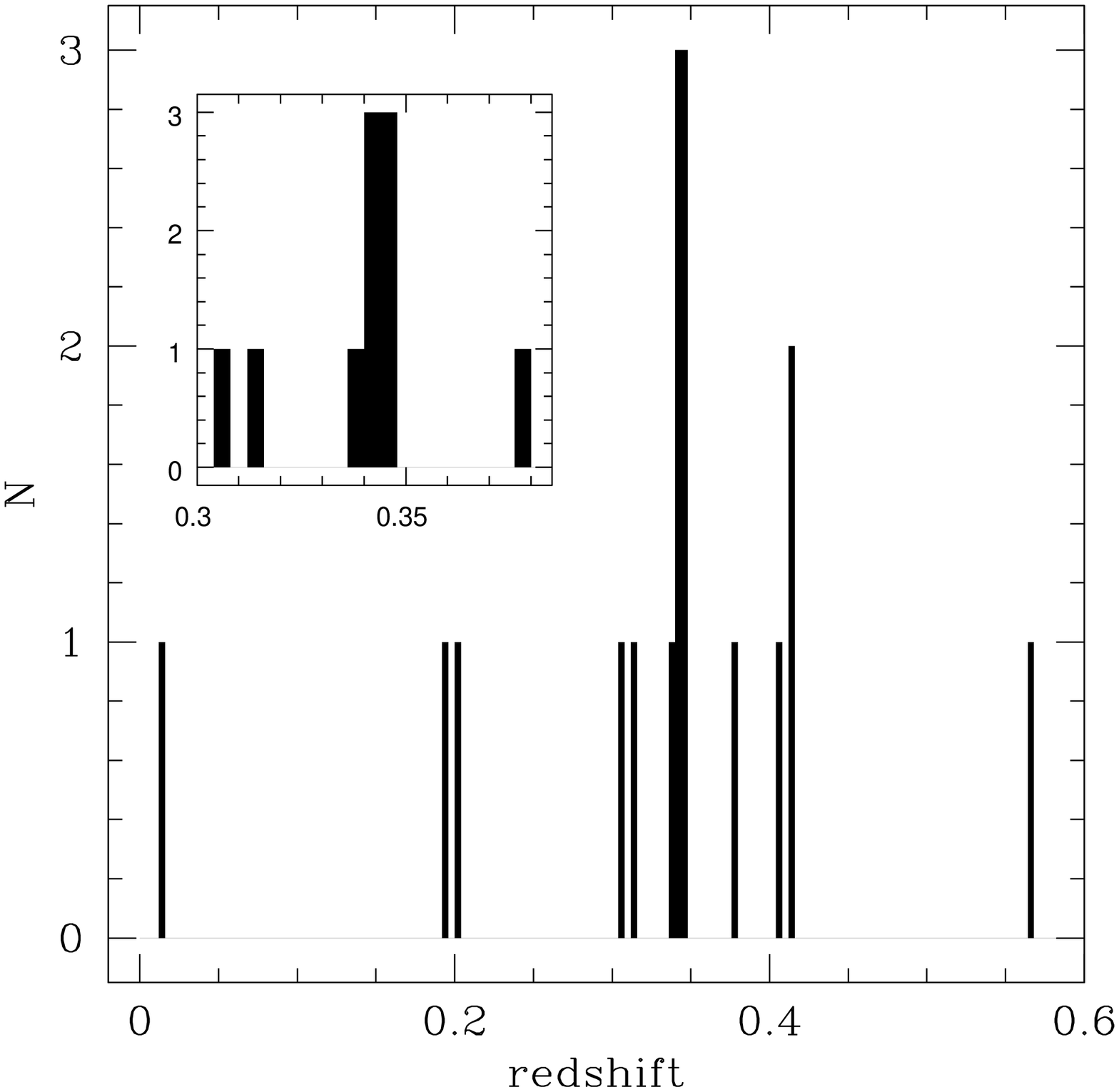}
   \caption{Redshift distribution of the observed galaxies in the line of sight of PKS 0447-439. 
The inset shows a narrower redshift range around the new group of galaxies.}
   \label{fig:z}
   \end{figure}

Figure \ref{fig:z} shows the redshift distribution of the 17 galaxies with measured redshift in the 
line of sight of PKS 0447-439. There is a group of galaxies at $z \sim 0.34$ 
consisting of seven galaxies. At this redshift, the field of view of GMOS is $1.45\times1.45$ Mpc. 
In Fig. \ref{fig:gemini}, the group members are marked with squares.
We computed the mean redshift ($z_{mean}$), the virial radius ($R_{vir}$) and the velocity dispersion 
($\sigma_v$) of the new group. The virial radius was computed following \citet{Nurmi:2013}
$${ R }_{ vir }=\left( \frac { 1 }{ { n }_{ p } } \sum { \frac { 1 }{ { r }_{ ij } }  }  \right) ^{ -1 },$$

\noindent where $r_{ij}$ is the projected distance between galaxies $i$ and $j$ 
for $n_{p}$ pairs. The mean redshift and velocity dispersion were calculated using the $Gapper$ 
estimator defined by \citet{Beers:1990}. We derived the following values: $z_{mean}=0.343$; 
$R_{vir}=0.42$ $\Mpc$ and 
$\sigma_v=622$ km/s, which correspond to a high-mass group of galaxies. We were unnot able to take spectra of several other galaxies in the field of view of  PKS 0447-439 
as a result of overlapping slits. Some of these galaxies show magnitudes that are comparable with those 
of the group members. Therefore, the number of seven members should be taken as a lower limit. 
There is a pair of galaxies at $z \sim 0.41$. 

\begin{table*}
\center
\begin{tabular}{rcccll}
\hline \hline

Slit  &R.A. (J2000.0)&Decl. (J2000.0)&$m_g'$&redshift/velocity         & comment \\
(1)   &   (2)     &  (3)         &  (4)  & (5)                         & (6) \\
\hline 
1  & 04:49:12.445 & -43:48:26.81 & 20.52 & z = 0.2023  $\pm$ 0.0002      & (a+e) merger  \\
2  & 04:49:15.279 & -43:48:18.28 & 20.85 & v = -112  $\pm$ 30 $\kms$   & (a) star \\
3  & 04:49:14.687 & -43:50:09.11 & 21.89 & z = 0.5677  $\pm$ 0.0001      & (a+e) \\
4  & 04:49:24.655 & -43:50:10.06 & 14.69 &                             & PKS 0447-439 \\
5  & 04:49:22.458 & -43:49:36.00 & 22.46 & z = 0.3463  $\pm$ 0.0002  [*] & (a) \\
6  & 04:49:21.540 & -43:50:32.75 & 22.15 & z = 0.3438  $\pm$ 0.0001  [*] & (e) spiral? \\ 
7  & 04:49:20.393 & -43:49:08.37 & 22.73 & z = 0.3055  $\pm$ 0.0002      & (a) \\
8  & 04:49:23.399 & -43:50:28.13 & 21.55 & z = 0.3458  $\pm$ 0.0001  [*] & (e) interacting \\
9  & 04:49:23.929 & -43:49:15.51 & 22.87 & z = 0.3153  $\pm$ 0.0005      & (a) interacting \\
10 & 04:49:19.931 & -43:50:33.49 & 20.28 & z = 0.1942  $\pm$ 0.0001      & (a+e) \\
11 & 04:49:30.571 & -43:51:44.91 & 18.70 & v = 303   $\pm$ 10 $\kms$   & (a) star \\
12 & 04:49:31.161 & -43:50:18.97 & 21.54 & z = 0.3401  $\pm$ 0.0001  [*] & (a) elongated-interacting \\
13 & 04:49:32.131 & -43:47:59.12 & 21.61 & z = 0.3406  $\pm$ 0.0001  [*] & (a) \\
14 & 04:49:28.429 & -43:52:22.24 & 21.81 & z = 0.3398  $\pm$ 0.0001  [*] & (a) \\
15 & 04:49:27.931 & -43:52:26.44 & 20.79 & z = 0.4155  $\pm$ 0.0001      & (a+e)  \\
16 & 04:49:33.070 & -43:49:27.06 & 22.21 & ?                           & (a+e) spiral-interacting \\
17 & 04:49:29.962 & -43:49:59.50 & 20.90 & z = 0.0156  $\pm$ 0.0001      & (e) elongated \\
18 & 04:49:26.109 & -43:50:45.69 & 22.35 & z = 0.4076  $\pm$ 0.0001      & (a+e) \\
19 & 04:49:27.256 & -43:49:17.23 & 22.17 & z = 0.3463  $\pm$ 0.0002  [*] & (a) \\
20 & 04:49:34.382 & -43:52:33.67 & 21.71 & z = 0.4133  $\pm$ 0.0001      & (a) \\
21 & 04:49:35.463 & -43:47:52.18 & 21.40 & z = 0.3761  $\pm$ 0.0002      & (e+a) spiral? \\  
22 & 04:49:36.763 & -43:49:36.69 & 21.78 & v = 150   $\pm$ 20 $\kms$   & (a) ? \\
\hline
\end{tabular}
\caption{Spectroscopically observed objects. Column 1 gives the slit numbers,
Cols. 2 and 3 show RA and Dec (J2000.0), Col. 4 shows the total $g'$ integrated magnitude,
Col. 5 lists the measured redshifts or radial velocity, and Col. 6 indicates whether the redshift was
computed using absorption (a), emission lines (e), or both (a+e). This
last column also gives a brief description of the observed object.}
\label{tab:data}
\end{table*}


\subsection{PKS 0447-439 as a group member}

As was previously discussed, BL-Lac objects are active nuclei hosted by massive elliptical galaxies. 
Since the former are usually found in groups or clusters of galaxies, it is natural 
to assume that PKS 0447-439 is a member of the group at $z = 0.343$. First of all, we tested the probability
of finding by chance a group of galaxies with seven or more members around a random position in the sky. 
\citet{Knobel:2009} identified  hundreds of groups in the redshift range $0.1 \lesssim z \lesssim 1.0$, 102 of them with
 5 or more member galaxies, in a $\sim$ 1.7 deg$^2$ of the COSMOS field \citep{Scoville:2007}. Using a 
Monte Carlo procedure, we
made catalogs of random positions in the same field of view as
that of the \citet{Knobel:2009} catalog.
Around each random position, we searched for coincidences with groups of galaxies  
in the \citet{Knobel:2009} catalog that would have seven or more members. The search was performed within a projected 
distance equal to the virial radius of the new group ($0.42$ Mpc). Since the \citet{Knobel:2009} catalog 
is deeper ($z \lesssim 1$) than our sample ($z \lesssim 0.6$), we restricted the analysis to the same redshift range. 
We found that the probability of finding 
by chance a group of galaxies in the redshift range $0 < z \lesssim 0.6$ around a random position is 9\%. 
If no restriction is imposed on the redshift range of the group catalog, the resulting probability 
increases by only 3\%.

If  the host galaxy of PKS 0447-439 is not a member of the reported group, it
probably is an isolated galaxy since no other structure is found in the line of sigh
of the BL-Lac. As discussed, BL-Lac are typically associated with
groups or clusters of galaxies. Consequently, the probability that the host galaxy of 
PKS 0447-439 is not a member of the reported group is the joint probability of having both 
an isolated host ($P_{ih}$) and finding by chance a group of galaxies in the line 
of sight of PKS 0447-439 ($P_{cg}=9\%$). Because of the lack of deep spectroscopic surveys of 
galaxies in fields around blazars, there is no accurate estimate of $P_{ih}$.
Based on a photometric study, \citet{Wurtz:1997} found that the clustering environments 
of BL-Lac objects are on average poor clusters. Although this method only relies 
on images and therefore is rather inefficient to detect groups of galaxies, two-thirds of 
the environment studies by \citet{Wurtz:1997} show positive net counts of galaxies after background subtraction.  Combining photometric and
spectroscopic data, \citet{Pesce:1995} found similar results. Based on this evidence, we consider 
that a conservative estimate of the probability to have a blazar in an isolated host is 
$P_{ih} \lesssim 0.3$. Since $P_{ih}$ and $P_{cg}$ are the probabilities of
two independent events, the joint probability of having both is 
$P_{ih} \times P_{cg} \lesssim 3\%$. In other words, the probability that the host galaxy of 
PKS 0447-439 is a member of the reported group is $\gtrsim 97\%$.

\citet{Landt:2002} suggested that BL-Lac spectra become featureless for jet-to-galaxy luminosity 
ratios $\gtrsim$10. In the case of PKS 0447-439, where the spectrum was taken with a high signal-to-noise ratio, 
a jet-to-galaxy ratio $\gtrsim$20-40 might be more 
appropriate (compare the spectrum in Fig. 1 of \citealt{Rovero:2013} with the simulated spectra in Fig. 1 of 
\citealt{Landt:2002}). Assuming that the host galaxy is a bright elliptical, we can estimate the expected 
jet-to-galaxy ratio assuming that PKS 0447-439 has a redshift $z=0.343$. For
an absolute magnitude $M_R \sim -23$ (a typical value for the host galaxy of BL-Lacs) we found a jet-to-galaxy 
ratio $\gtrsim$30. This high value is consistent with the 
featureless spectrum of PKS 0447-439. 

\begin{figure}
   \centering
   \includegraphics[width=8cm]{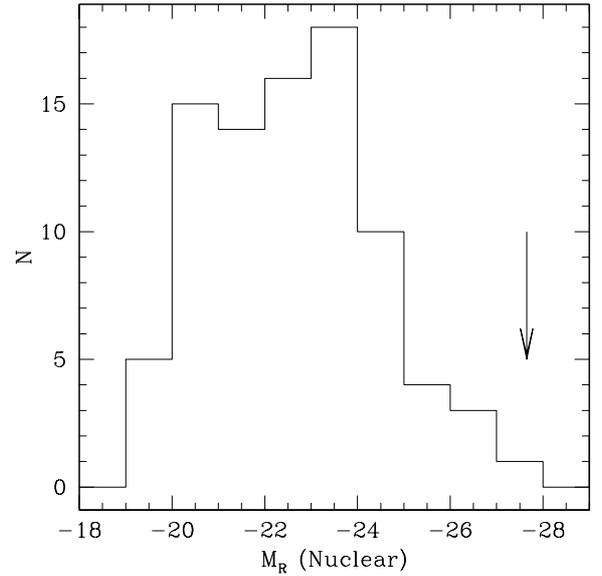}
   \caption{Distribution of the absolute R-band jet magnitude of the BL-Lac objects in the XBL catalog. 
The vertical arrow shows the absolute magnitude of PKS 0447-439 at a redshift of 0.343.}
   \label{fig:xbl}
   \end{figure}

Assuming a $z=0.343$ for PKS 0447-439 and the apparent magnitude $m_R=13.58$ obtained in Sect. \ref{sec:bosque}, 
we estimated an absolute magnitude of $M_R=-27.65$, which was computed as
$${ M }_{ R }={ m }_{ R }-{ A }_{ R }+K(z)+E(z)-5\log { { d }_{ L } } +5,$$

\noindent where $d_L$ is the luminosity distance; $A_R$ is the galactic reddening; $K(z)$ is the 
$K$-correction, and $E(z)$ is the 
evolutionary correction. $K(z)$  and $E(z)$ were computed following \citet{Fan:2000}.
The derived value can be compared with those from other BL-Lacs. 
\citet{Kapanadze:2013} compiled a sample of X-ray selected
BL-Lacs (XBL catalog); for 94 out of 312 objects they quoted both the host and the jet contributions to 
the total luminosity. Figure \ref{fig:xbl} shows the absolute R-band jet magnitude of BL-Lac objects 
in the XBL catalog. The vertical arrow corresponds to the absolute magnitude of PKS 0447-439. 
Note that this comparison neglects the host contribution. Clearly, PKS 0447-439 
is among the brightest BL-Lacs, which is also consistent with the featureless observed spectra. 
If the blazar were at a much higher redshift, the jet luminosity of PKS 0447-439 would have an 
unprecedentedly high brightness. On the other hand, if the blazar were at a much lower redshift, host galaxy contribution 
should be observed. Based on this idea, \citet{Landt:2008} estimated for PKS 0447-439 a lower limit of 
z > 0.176 (assuming a jet-to-galaxy ratio of ten).

\section{Discussion and conclusions}
\label{sec:Con}

Using the Gemini Multi-Object Spectrograph, we obtained the spectra of twenty-one objects in the field of 
view of the blazar PKS 0447-439. At least 17 of these objects are galaxies in the redshift range
$0.015 \lesssim  z \lesssim 0.56$. We identified a group of galaxies at $z=0.343$ with seven
members. The group has a virial radius of $R_{vir}= 0.42$ $\Mpc$ and a velocity dispersion 
$\sigma =622$ $\kms$. 
We estimated a $9\%$ probability of finding by chance a group 
like the one observed. 
If  the host galaxy of PKS 0447-439 is not a member of the reported group, it
probably is an isolated galaxy since no other structure is found in the line of sight
of the BL Lac. We found that the joint probability of having both,
an isolated host  and finding by chance a group of galaxies in the line 
of sight of PKS 0447-439 is $\lesssim 3\%$. Therefore, the probability that the host galaxy of 
PKS 0447-439 is a member of the reported group is $\gtrsim 97\%$.
Considering these results, we 
propose to adopt $z \sim 0.343 \pm 0.002$ as the most likely redshift for PKS 0447-439.  
The quoted uncertainty corresponds to the velocity dispersion of the parent group.

For a redshift of 0.343, we derived an absolute magnitude of $M_R=-27.65$. Because the 
spectrum of PKS 0447-439 is featureless, we assume that the observed luminosity is mostly 
due to the jet contribution. Under this assumption, PKS 0447-439 is among the brightest 
BL-Lacs. We also estimated a lower limit for the jet-to-host ratio assuming that the host galaxy is a 
bright elliptical and found a jet-to-galaxy ratio $\gtrsim$20-40. For this high ratio, a featureless 
spectrum is expected \citep{Landt:2002}. Therefore, the observed spectrum of PKS 0447-439 is also 
consistent with the proposed redshift. 

From the point of view of gamma-ray astronomy, the redshift proposed here agrees with other 
estimates in the literature. 
However, it has to be stressed that all these estimates used 
some level of assumption on one or more of the variables involved in the estimation, such as the EBL 
density and the intrinsic spectrum shape. On the other hand, redshift estimates from spectroscopic 
observations have failed to determine the value for PKS 0447-439, either because they found no 
features at all or because the lines used were very weak and the measurement was never confirmed by 
independent observations. Perhaps the spectroscopic procedure using spectra at far-ultraviolet 
\citep{Danforth:2010, Danforth:2013, Furniss:2013} could be used to confirm the redshift 
of this blazar, for which HST/COS observations need to be performed.
We here provided a totally independent estimate for 
the redshift of PKS 0447-439. The method applied in this work to estimate the redshift of PKS 0447-439
can be applied to other BL-Lacs with featureless spectra, for which deep spectroscopic observations 
would be necessary. 

\begin{acknowledgements}
This work is based on observations obtained at the Gemini Observatory, which is operated by the 
Association of Universities for Research in Astronomy, Inc., under a cooperative agreement with 
the NSF on behalf of the Gemini partnership: the National Science Foundation (United States), 
the National Research Council (Canada), CONICYT (Chile), the Australian Research Council 
(Australia), Minist\'{e}rio da Ci\^{e}ncia, Tecnologia e Inova\c{c}\~{a}o (Brazil) and Ministerio 
de Ciencia, Tecnolog\'{i}a e Innovaci\'{o}n Productiva (Argentina).
This work has been partially supported with grants from Consejo Nacional de Investigaciones 
Cient\'{i}ficas y T\'{e}cnicas de la Rep\'{u}blica Argentina (CONICET) and Secretar\'{i}a de 
Ciencia y Tecnolog\'{i}a de la Universidad de C\'ordoba. The authors H.M., C.D., and A.C.R. 
are members of ``Carrera del Investigador Cient\'ifico'' of CONICET, Argentina, and A.P. is a 
postdoctoral fellow of CONICET, Argentina.
\end{acknowledgements}


\bibliographystyle{aa} 
\bibliography{25050} 

\end{document}